# Secure and Privacy-Preserving Authentication Protocols for Wireless Mesh Networks

Jaydip Sen

*Innovation Lab, Tata Consultancy Services Ltd.*
*India*

## 1. Introduction

*Wireless mesh networks* (WMNs) have emerged as a promising concept to meet the challenges in next-generation wireless networks such as providing flexible, adaptive, and reconfigurable architecture while offering cost-effective solutions to service providers (Akyildiz et al., 2005). WMNs are multi-hop networks consisting of *mesh routers* (MRs), which form wireless mesh backbones and *mesh clients* (MCs). The mesh routers provide a rich radio mesh connectivity which significantly reduces the up-front deployment cost of the network. Mesh routers are typically stationary and do not have power constraints. However, the clients are mobile and energy-constrained. Some mesh routers are designated as gateway routers which are connected to the Internet through a wired backbone. A gateway router provides access to conventional clients and interconnects ad hoc, sensor, cellular, and other networks to the Internet. The gateway routers are also referred to as the *Internet gateways* (IGWs). A mesh network can provide multi-hop communication paths between wireless clients, thereby serving as a community network, or can provide multi-hop paths between the client and the gateway router, thereby providing broadband Internet access to the clients.

As WMNs become an increasingly popular replacement technology for last-mile connectivity to the home networking, community and neighborhood networking, it is imperative to design efficient and secure communication protocols for these networks. However, several vulnerabilities exist in the current protocols of WMNs. These security loopholes can be exploited by potential attackers to launch attack on WMNs. Absence of a central point of administration makes securing WMNs even more challenging. Security is, therefore, an issue which is of prime importance in WMNs (Sen, 2011). Since in a WMN, traffic from the end users is relayed via multiple wireless mesh routers, preserving privacy of the user data is also a critical requirement (Wu et al., 2006a). Some of the existing security and privacy protection protocols for WMNs are based on the trust and reputation of the network entities (Sen, 2010a; Sen, 2010b). However, many of these schemes are primarily designed for *mobile ad hoc networks* (MANETs) (Sen, 2006; Sen, 2010c), and hence these protocols do not perform well in large-scale hybrid WMN environments.

The broadcast nature of transmission and the dependency on the intermediate nodes for multi-hop communications lead to several security vulnerabilities in WMNs. The attacks can be external as well as internal in nature. External attacks are launched by intruders who are not authorized users of the network. For example, an intruding node may eavesdrop on the packets and replay those packets at a later point of time to gain access to the network resources. On the other hand, the internal attacks are launched by the nodes that are part of the WMN. On example of such attack is an intermediate node dropping packets which it was supposed to forward. To prevent external attacks in vulnerable networks such as WMNs, strong authentication and access control mechanisms should be in place for practical deployment and use of WMNs. A secure authentication should enable two communicating entities (either a pair of MC and MR or a pair of MCs) to validate the authenticity of each other and generate the shared common session keys which can be used in cryptographic algorithms for enforcing message confidentiality and integrity. As in other wireless networks, a weak authentication scheme can easily be compromised due to several reasons such as distributed network architecture, the broadcast nature of the wireless medium, and dynamic network topology (Akyildiz et al., 2005). Moreover, the behavior of an MC or MR can be easily monitored or traced in a WMN by adversaries due to the use of wireless channel, multi-hop connection through third parties, and converged traffic pattern traversing through the IGW nodes. Under such scenario, it is imperative to hide an active node that connects to an IGW by making it anonymous. Since on the Internet side traditional anonymous routing approaches are not implemented, or may be compromised by strong attackers such protections are extremely critical (X. Wu & Li, 2006).

This chapter presents a comprehensive discussion on the current authentication and privacy protection schemes for WMN. In addition, it proposes a novel security protocol for node authentication and message confidentiality and an anonymization scheme for privacy protection of users in WMNs.

The rest of this chapter is organized as follows. Section 2 discusses the issues related to access control and authentication in WMNs. Various security vulnerabilities in the authentication and access control mechanisms for WMNs are first presented and then a list of requirements (i.e. properties) of a secure authentication scheme in an open and large-scale, hybrid WMN are discussed. Section 3 highlights the importance of the protection user privacy in WMNs. Section 4 presents a state of the art survey on the current authentication and privacy protection schemes for WMNs. Each of the schemes is discussed with respect to its applicability, performance efficiency and shortcomings. Section 5 presents the details of a hierarchical architecture of a WMN and the assumptions made for the design of a secure and anonymous authentication protocol for WMNs. Section 6 describes the proposed key management scheme for secure authentication. Section 7 discusses the proposed privacy protection algorithm which ensures user anonymity. Section 8 presents some performance results of the proposed scheme. Section 9 concludes the chapter while highlighting some future direction of research in the field of secure authentication in WMNs.

## 2. Access control and authentication in WMNs

Authentication and authorization is the first step towards prevention of fraudulent accesses by unauthorized users in a network. Authentication ensures that an MC and the corresponding MR can mutually validate their credentials with each other before the MC is allowed to access the network services. In this section, we first present various attacks in WMNs that can be launched on the authentication services and then enumerate the requirements for authentication under various scenarios.

## 2.1 Security vulnerabilities in authentication schemes

Several vulnerabilities exist in different protocols for WMNs. These vulnerabilities can be suitably exploited by potential attackers to degrade the network performance (Sen, 2011). The nodes in a WMN depend on the cooperation of other nodes in the network for their successful operations. Consequently, the *medium access control* (MAC) layer and the network layer protocols for these networks usually assume that the participating nodes are honest and well-behaving with no malicious or dishonest intentions. In practice, however, some nodes in a WMN may behave in a selfish manner or may be compromised by malicious users. The assumed trust (which in reality may not exist) and the lack of accountability due to the absence of a central point of administration make the MAC and the network layer protocols vulnerable to various types of attacks. In this sub-section, we present a comprehensive discussion on various types of attacks on the existing authentication schemes of WMNs. A detailed list various attacks on the different layers of WMN communication protocol stack can be found in (Sen, 2011; Yi et al., 2010).

There are several types of attacks that are related to authentication in WMNs. These attacks are: (i) unauthorized access, (ii) replay attack, (iii) spoofing attack, (iv) denial of service attack (DoS), (v) intentional collision of frames, (vi) pre-computation and partial matching attack, and (vi) compromised or forged MRs. These attacks are discussed in detail below.

**Unauthorized access**: in this attack, an unauthorized user gets access to the network services by masquerading a legitimate user.

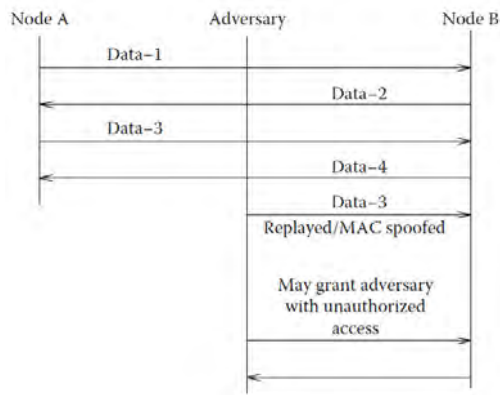

Fig. 1. Illustration of MAC spoofing and replay attacks [Source: (Sen, 2011)]

**Replay attack:** the replay attack is a type of *man-in-the-middle* attack (Mishra & Arbaugh, 2002) that can be launched by external as well as internal nodes. An external malicious node can eavesdrop on the broadcast communication between two nodes (*A* and *B*) in the network as shown in Fig. 1. It can then transmit legitimate messages at a later point of time to gain access to the network resources. Generally, the authentication information is replayed where the attacker deceives a node (node *B* in Fig. 1) to believe that the attacker is a legitimate node (node *A* in Fig. 1). On a similar note, an internal malicious node, which is an intermediate hop between two communicating nodes, can keep a copy of all relayed data. It can then retransmit this data at a later point in time to gain unauthorized access to the network resources.

**Spoof attack:** spoofing is the act of forging a legitimate MAC or IP address. IP spoofing is quite common in multi-hop communications in WMNs. In IP spoofing attack, an adversary inserts a false source address (or the address of a legitimate node) from the packets forwarded by it. Using such a spoofed address, the malicious attacker can intercept a termination request and hijack a session. In MAC address spoofing, the attacker modifies the MAC address in transmitted frames from a legitimate node. MAC address spoofing enables the attacker to evade *intrusion detection systems* (IDSs) that may be in place.

**DoS attack:** in this attack, a malicious attacker sends a flood of packets to an MR thereby making a buffer overflow in the router. Another well-known security flaw can be exploited by an attacker. In this attack, a malicious attacker can send false termination messages on behalf of a legitimate MC thereby preventing a legitimate user from accessing network services.

**Intentional collision of frames:** a collision occurs when two nodes attempt to transmit on the same frequency simultaneously (Wood & Stankovic, 2002). When frames collide, they are discarded and need to be retransmitted. An adversary may strategically cause collisions in specific packets such as acknowledgment (ACK) control messages. A possible result of such collision is the costly exponential back-off. The adversary may simply violate the communication protocol and continuously transmit messages in an attempt to generate collisions. Repeated collisions can also be used by an attacker to cause resource exhaustion. For example, a naïve MAC layer implementation may continuously attempt to retransmit the corrupted packets. Unless these retransmissions are detected early, the energy levels of the nodes would be exhausted quickly. An attacker may cause unfairness by intermittently using the MAC layer attacks. In this case, the adversary causes degradation of real-time applications running on other nodes by intermittently disrupting their frame transmissions.

**Pre-computation and partial matching attack:** unlike the attacks mentioned above, where the MAC protocol vulnerabilities are exploited, these attacks exploit the vulnerabilities in the security mechanisms that are employed to secure the MAC layer of the network. Pre-computation and partial matching attacks exploit the cryptographic primitives that are used at the MAC layer to secure the communication. In a pre-computation attack, or *time memory trade-off* (TMTO) attack, the attacker computes a large amount of information (e.g., key, plaintext, and the corresponding ciphertext) and stores that information before launching the attack. When the actual transmission starts, the attacker uses the pre-computed information to speed up the cryptanalysis process. TMTO attacks are highly effective against a large number of cryptographic solutions. On the other hand, in a partial matching attack, the attacker has access to some (ciphertext, plaintext) pairs, which in turn decreases the encryption key strength, and improves the chances of success of the brute force mechanisms. Partial matching attacks exploit the weak implementations of encryption algorithms. For example, the IEEE 802.11i standard for MAC layer security in wireless networks is prone to the session hijacking attack and the *man-in-the-middle* attack that exploits the vulnerabilities in IEEE802.1X. DoS attacks are possible on the four-way handshake procedure in IEEE802.11i.

**Compromised or Forged MR:** an attacker may be able to compromise one or more MRs in a network by physical tampering or logical break-in. The adversary may also introduce rogue MRs to launch various types of attacks. The fake or compromised MRs may be used to attack the wireless link thereby implementing attacks such as: passive eavesdropping, jamming, replay and false message injection, traffic analysis etc. The attacker may also advertise itself as a genuine MR by forging duplicate beacons procured by eavesdropping on genuine MRs in the network. When an MC receives these beacon messages, it assumes that it is within the radio coverage of a genuine MR, and initiates a registration procedure. The false MR now can extract the secret credentials of the MC and can launch spoof attack on the network. This attack is possible in protocols which require an MC to be authenticated by and MR but not the vice versa (He et al., 2011).

## 2.2 Requirements for authentication in WMNs

On the basis of whether a central authentication server is available, there are two types of implementations of access control enforcements in WMNs: (i) centralized access control and (ii) distributed access control. For both these approaches, the access control policies should be implemented at the border of the mesh network. In the distributed access control, the access points could act as the distributed authentication servers. The authentication could also be performed in three different places:

- A remote central authentication center
- Local entities such as IGWs or MRs  that play the role of an authentication server
- Local MRs

The main benefit of central authentication server is the ease of management and maintenance. However, this approach suffers from the drawback of having a single point of failure. Due to higher *round trip time* (RTT) and authentication delay, a centralized authentication scheme in a multi-hop WMN is not desirable. Instead, authentication protocols are implemented in local nodes such as IGW or MRs. For ensuring higher level of availability of the network services, the authentication power is delegated to a group of MRs in order to avoid single point of failure.

The objective of an authentication system is to guarantee that only the legitimate users have access to the network services. Any pair of network entities in a WMN (e.g., IGW, MR, and MC) may need to mutually authenticate if required. An MR and MC should be able to mutually authenticate each other to prevent unauthorized network access and other attacks. The MCs and MRs should be able to establish a shared pair-wise session key to encrypt messages. The protocol should have robust key generation, distribution and revocation procedures.

Several requirements have been identified in (Buttyan et al., 2010) for authentication mechanisms between MC and MRs in a WMN. These requirements are summarized below:

- *Authentication should be fast enough to support user mobility*. In order to maintain the *quality of service* (QoS) of user applications on mobile MCs, the authentication process should be fast. Also, the re-authentication delays should be within the acceptable limit of handoff delay.
- *MCs and MRs should be able to authenticate themselves mutually*. During the authentication process, the MR authenticates the MC, but the MR also should prove its authenticity to the MC.
- *Authentication process should be resistant to DoS attacks*. Since a successful attack against the central authentication server will lead to a complete compromise of the security system in the network, the authentication process should be robust.

- *Authentication protocols should be compatible with standards*. In a multi-operator environment, it is mandatory that the authentication protocols are standardized so that an MC of one vendor should be able to authenticate with the MR of a different network operator.
- *Authentication protocols should be scalable*. Since the mesh networks have large number of MCs, MRs and IGWs, the authentication protocol should be scalable and must not degrade in performance as the network size increases.

The mutual authentication protocols for MCs and MRs must use several keys for encrypting the credentials. The connection key management should satisfy the following requirements.

- *The connection keys should not reveal long term keys*. The connection keys that the MRs obtain during the authentication of the MCs should not reveal any long-term authentication keys. This requirement must hold because in the multi-operator environment, the MCs may associate to MRs operated by foreign operators.
- *The connection keys should be independent of each other*. As the neighboring MRs may not fully trust each other in a multi-operator environment, the authentication and key generation mechanism have to prevent an MR from deriving connection keys that are used at another MR.
- *The connection keys must be fresh in each session*. It must be ensured that the connection key derived during the authentication protocol for both participants (MC and MR) is fresh.

## 3. User privacy requirement in WMNs

Privacy provision is an important issue to be considered for WMN deployment. However, privacy is difficult to achieve even if messages are protected, as there are no security solutions or mechanisms which can guarantee that data is not revealed by the authorized parties themselves (Moustafa, 2007). Thus, it is important that complementary solutions are in place. Moreover, communication privacy cannot not be assured with message encryption since the attackers can still observe who is communicating with whom as well as the frequency and duration of the communication sessions. This makes personal information susceptible to disclosure and subsequent misuse even when encryption mechanisms are in place. Furthermore, users in WMNs can be easily monitored or traced with regard to their presence and location, which causes the exposure of their personal life. Unauthorized parties can get access to the location information about the MC's positions by observing their communications and traffic patterns. Consequently, there is a need to ensure location privacy in WMNs as well.

To control the usage of personal information and the disclosure of personal data, different types of information hiding mechanisms like anonymity, data masking etc should be implemented in WMN applications. The following approaches can be useful in information hiding, depending on what is needed to be protected:

- *Anonymity*: this is concerned with hiding the identity of the sender or receiver of the message or both of them. In fact, hiding the identity of both the sender and the receiver of the message can assure communication privacy. Thus, attackers monitoring the messages being communicated could not know who is communicating with whom, thus no personal information is disclosed.

- *Confidentiality*: it is concerned with hiding the transferred messages by using suitable data encryption algorithms. Instead of hiding the identity of the sender and the receiver of a message, the message itself is hidden in this approach.
- *Use of pseudonyms*: this is concerned with replacing the identity of the sender and the receiver of the message by pseudonyms which function as identifiers. The pseudonyms can be used as a reference to the communicating parties without infringing on their privacy, which helps to ensure that the users in the WMNs cannot be traced or identified by malicious adversaries. However, it is important to ensure that there exist no indirect ways by which the adversaries can link the pseudonyms with their corresponding real world entities.

Privacy has been a major concern of Internet users (Clarke, 1999). It is also been a particularly critical issue in context of WMN-based Internet access, where users' traffic is forwarded via multiple MRs. In a community mesh network, this implies that the traffic of a residence can be observed by the MRs residing at its neighbors premises. Therefore, privacy in WMNs has two different dimensions: (i) data confidentiality (or privacy) and traffic confidentiality.  These issues are briefly described below:

- *Data confidentiality*: it is obvious that data content reveals user privacy on what is being communicated. Data confidentiality aims to protect the data content and prevent eavesdropping by intermediate MRs. Message encryption is a conventional approach for data confidentiality.
- *Traffic confidentiality*: traffic information such as with whom, when and how frequently the users are communicating, and the pattern of traffic also reveal critical privacy-sensitive information. The broadcast nature of wireless communication makes acquiring such information easy. In a WMN, attackers can conduct traffic analysis as MRs by simply listening to the channels to identify the "ups and downs" of the target's traffic. While data confidentiality can be achieved via message encryption, it is much harder to preserve traffic confidentiality (T. Wu et al., 2006).

## 4. Secure authentication and privacy protection schemes in WMNs

Since security and privacy are two extremely important issues in any communication network, researchers have worked on these two areas extensively. However, as compared to MANETs and *wireless sensor networks* (WSNs) (Sen, 2009; Sen & Subramanyam, 2007), WMNs have received very little attention in this regard. In this section, we first present a brief discussion on some of the existing propositions for secure authentication and user privacy protection in WMNs. Later on, some of the mechanisms are discussed in detail in the following sub-sections.

In (Mishra & Arbaugh, 2002), a standard mechanism has been proposed for client authentication and access control to guarantee a high-level of flexibility and transparency to all users in a wireless network. The users can access the mesh network without requiring any change in their devices and softwares. However, client mobility can pose severe problems to the security architecture, especially when real-time traffic is transmitted. To cope with this problem, *proactive key distribution* has been proposed in (Kassab et al., 2005; Prasad & Wang, 2005).

Providing security in the backbone network for WMNs is another important challenge. Mesh networks typically employ resource constrained mobile clients, which are difficult to protect against removal, tampering, or replication. If the device can be remotely managed, a distant hacking into the device would work perfectly (Ben Salem & Hubaux, 2006). Accordingly, several research works have been done to investigate the use of cryptographic techniques to achieve secure communication in WMNs. In (Cheikhrouhou et al., 2006), a security architecture has been proposed that is suitable for multi-hop WMNs employing PANA (Protocol for carrying Authentication for Network Access) (Parthasarathy, 2006). In the scheme, the wireless clients are authenticated on production of the cryptographic credentials necessary to create an encrypted tunnel with the remote access router to which they are associated. Even though such framework protects the confidentiality of the information exchanged, it cannot prevent adversaries to perform active attacks against the network itself. For instance, a malicious adversary can replicate, modify and forge the topology information exchanged among mesh devices, in order to launch a denial of service attack. Moreover, PANA necessitates the existence of IP addresses in all the mesh nodes, which is poses a serious constraint on deployment of this protocol.

Authenticating transmitted data packets is an approach for preventing unauthorized nodes to access the resources of a WMN. A *light-weight hop-by-hop access protocol* (LHAP) has been proposed for authenticating mobile clients in wireless dynamic environments, preventing resource consumption attacks (Zhu et al., 2006). LHAP implements light-weight hop-by-hop authentication, where intermediate nodes authenticate all the packets they receive before forwarding them. LHAP employs a packet authentication technique based on the use of one-way hash chains. Moreover, LHAP uses TESLA (Perrig et al., 2001) protocol to reduce the number of public key operations for bootstrapping and maintaining trust between nodes.

In (Prasad et al., 2004), a lightweight *authentication, authorization and accounting* (AAA) infrastructure is proposed for providing continuous, on-demand, end-to-end security in heterogeneous networks including WMNs. The notion of a security manager is used through employing an AAA broker. The broker acts as a settlement agent, providing security and a central point of contact for many service providers.

The issue of user privacy in WMNs has also attracted the attention of the research community. In (T. Wu et al., 2006), a light-weight privacy preserving solution is presented to achieve well-maintained balance between network performance and traffic privacy preservation. At the center of the solution is of information-theoretic metric called *traffic entropy*, which quantifies the amount of information required to describe the traffic pattern and to characterize the performance of traffic privacy preservation. The authors have also presented a penalty-based shortest path routing algorithm that maximally preserves traffic privacy by minimizing the mutual information of traffic entropy observed at each individual relaying node while controlling the possible degradation of network within an acceptable region. Extensive simulation study proves the soundness of the solution and its resilience to cases when two malicious observers collude. However, one of the major problems of the solution is that the algorithm is evaluated in a single-radio, single channel WMN. Performance of the algorithm in multiple radios, multiple channels scenario will be a really questionable issue. Moreover, the solution has a scalability problem. In (X. Wu & Li, 2006), a mechanism is proposed with the objective of hiding an active node that connects to a gateway router, where the active mesh node has to be anonymous. A novel communication protocol is designed to protect the node's privacy using both cryptography and redundancy. This protocol uses the concept of *onion routing* (Reed et al., 1998). A mobile user who requires anonymous communication sends a request to an *onion router* (OR). The OR acts as a proxy to the mobile user and constructs an onion route consisting of other ORs using the public keys of the routers. The onion is constructed such that the inner most part is the message for the intended destination, and the message is wrapped by being encrypted using the public keys of the ORs in the route. The mechanism protects the routing information from insider and outsider attack. However, it has a high computation and communication overhead.

In the following sub-sections, some of the well-known authentication and privacy preservation schemes for WMNs are discussed briefly. For each of the schemes, its salient features and potential shortcomings are highlighted.

## 4.1 Local authentication based on public key certificates

In the localized authentication, a *trusted third party* (TTP) serves as the trusted *certificate authority* (CA) that issues certificates. In (Buttyan & Dora, 2009), a localized authentication scheme is proposed in which authentication is performed locally between the MCs and the MRs in a hybrid large-scale WMN operated by a number of operators. Each operator maintains its own CA. Each CA is responsible for issuing certificates to its customers. Each CA maintains its own *certificate revocation list* (CRL). The CAs also issue cross-certificates among each other for enabling entities (MCs or MRs) subscribing to different operators to perform certificate-based authentications and key exchanges. To minimize authentication delay, the *provably secure key transport protocol* (Blake-Wilson & Menezes, 1998) proposed by Blake-Wilson-Menezes (BWM) has been used.

For authentication in multiple domains in a metropolitan area network, a localized authentication scheme has been proposed in (Lin et al., 2008). In this scheme, an *embedded two-factor authentication* mechanism is utilized to verify the authenticity of a roaming MC. The authenticity verification does not need any intervention of the home *Internet service provider* (ISP) of the MC. The two-factor authentication mechanism includes two methods of authentication: password and smart card. To minimize the *ping-pong effect*, the session key is cached in the current network domain. Whenever the MC requests a handoff into a neighboring MR which has a valid shared session key with the MC, a user-authenticated key agreement protocol with secret key cryptography is performed. Thus an expensive full authentication based on an asymmetric key encryption is avoided. The protocol execution is fast since it involves encryption using only the symmetric key and keyed *hash message authentication codes* (HMACs).

The localized authentication schemes are based on the assumption that the MRs are trusted and fully protected by robust certificates. In practice, MRs are low cost devices and without extra protection, these devices can easily be compromised. In the event an MR gets compromised, the local authentication schemes will fail. To defend against compromised MRs, a scheme based on local voting strategy (Zhu et al., 2008) is adopted which work on the principle of *threshold digital signature* mechanism (Cao et al., 2006).

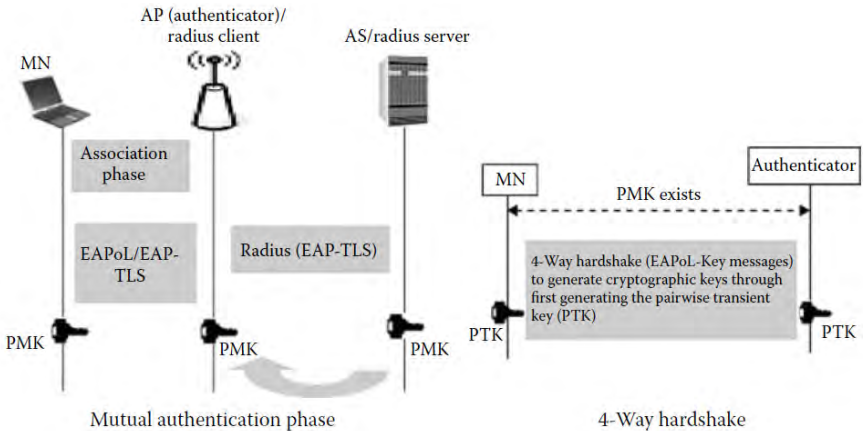

Fig. 2. Schematic diagram of IEEE 802.11i authentication protocol [Source: (Moustafa, 2007)]

## 4.2 Authentication model based on 802.11i protocol

In most commercial deployments of wireless local area networks (WLANs), IEEE 802.11i (IEEE 802.11i, 2004) is the most common approach for assuring authentication at the layer 2. However, the IEEE 802.11i authentication does not fully address the problem of WLAN vulnerability (Moustafa, 2007). In IEEE 802.11i authentication, as described in Fig. 2, the MC and the *authentication server* (AS) apply the 802.1X (IEEE 802.1X, 2001) authentication model carrying out some negotiation to agree on *pair-wise master key* (PMK) by using some upper layer authentication schemes or using a pre-shared secret. This key is generated by both the MC and the AS, assuring the mutual authentication between them. The *access point* (AP) then receives a PMK copy from the AS, authenticating the MC and authorizing its communication. Afterwards, a four-way handshake starts between the AP and the MC to generate encryption keys from the generated PMK. Encryption keys can assure confidential transfer between the MC and the AP. If the MC roams to a new AP, it will perform another full 802.1X authentication with the AS to derive a new PMK. For performance enhancement, the PMK of the MC is cached by the MC and the AP to be used for later re-association without another full authentication. The features of 802.11i exhibit a potential vulnerability because a compromised AP can still authenticate itself to an MC and gain control over the connection. Furthermore, IEEE 802.11i authentication does not provide a solution for multi-hop communication. Consequently new mechanisms are needed for authentication and secure layer 2 links setup in WMNs (Moustafa, 2007).

*Wireless dual authentication protocol* (WDAP) (Zheng et al., 2005) is proposed for 802.11 WLAN and can be extended to WMNs. WDAP provides authentication for both MCs and APs and overcomes the shortcomings of other authentication protocols. The name "dual" implies the fact that the AS authenticates both the MC and the AP. As in the four-way handshake in IEEE 802.11i, this protocol also generates a session key for maintaining confidentiality of the messages communicated between the MC and the AP after a successful authentication. WDAP provides authentication during the initial connection state. For roaming, it has three sub-protocols: an authentication protocol, a de-authentication protocol, and a roaming authentication protocol.

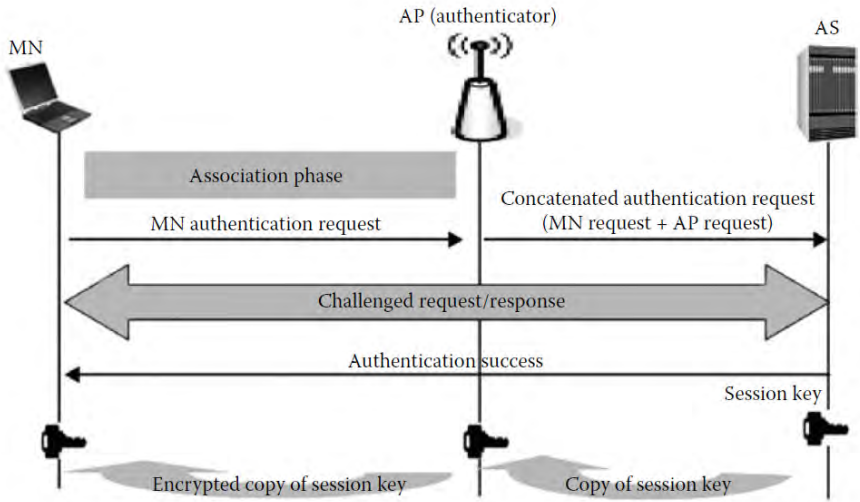

Fig. 3. Schematic diagram of the authentication process in WDAP [Source: (Moustafa, 2007)]

Fig. 3 illustrates the WDAP authentication process. In the authentication protocol, the AP receives the authentication request from the MC. It then creates an authentication request for itself and concatenates this request to the received request from the MC. The concatenated request is then sent to the AS. Since both the mobile station and the AP do not trust each other until the AS authenticates both of them, WDAP is a dual authentication protocol. If the authentication is successful, AS generates a session key and sends the key to the AP. The AP then sends this key to the MC encrypting it with the shared key with MC. This key is thus shared between the AP and the MC for their secure communication and secure de-authentication when the session is finished. When an MC finishes a session with an AP, secure de-authentication takes place to prevent the connection from being exploited by an adversary. Use of WDAP in WMN environments ensures mutual authentication of both MCs and MRs. Also, WDAP can be used to ensure authentication between the MRs through authentication requests concatenation. In case of multi-hop communication in WMNs, each pair of nodes can mutually authenticate through the session key generated by the AS. However, a solution is needed in case of open mesh networks scenarios, where the AS may not be present in reality. Another problem arises in case of roaming authentication. WDAP is not ideally suited for use in roaming authentication since it works only for roaming into new APs, and does not consider the case of *back roaming* in which an MC may need to re-connect with another MC or an AP with whom it was authenticated earlier. As a result, the WDAP session key revocation mechanisms has some shortcomings that makes it unsuitable for deployment in real-world WMNs.

An approach that adapts IEEE 802.11i to the multi-hop communication has been presented in (Moustafa et al., 2006a). An extended forwarding capability in 802.11i is proposed without compromising on its security features to setup authenticated links in layer 2 to achieve secure wireless access as well as confidential data transfer in ad hoc multi-hop environments. The general objective of this approach is to support secure and seamless access to the Internet by the MCs situated near public WLAN hotspots, even when these nodes may move beyond the coverage area of the WLAN. To accomplish the *authentication, authorization and accounting* (AAA) process for an MC within the WLAN communication range, classical 802.11i authentication and message exchange take place.

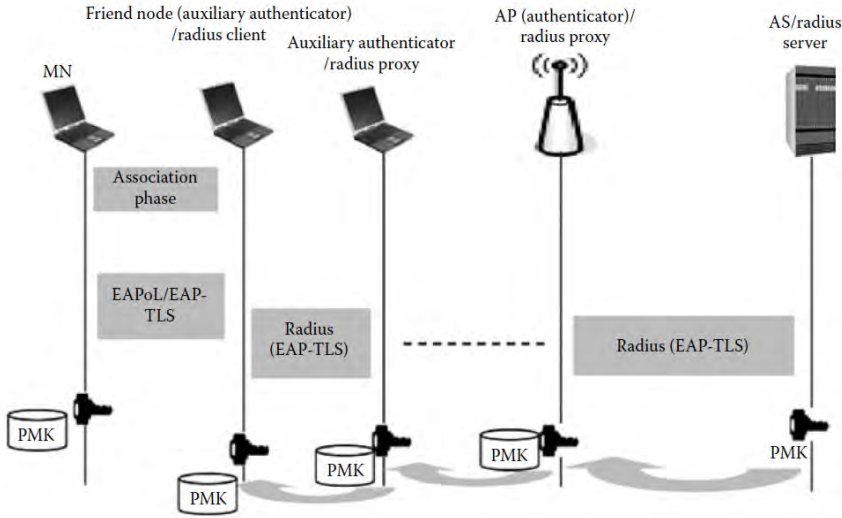

Fig. 4. Schematic diagram of adapted 802.11i with EAP-TLS for multi-hop communication [Source: (Moustafa, 2007)]

As shown in Fig. 4, for accomplishing the AAA process for MCs that are beyond the WLAN communication range but belong to the ad hoc clusters, 802.11i is extended to support forwarding capabilities. In this case, the notion of *friend nodes* is introduced to allow each MC to initiate the authentication process through a selected node in its proximity. The friend node plays the role of an auxiliary authenticator that forwards the authentication request of the MC to the actual authenticator (i.e., the AP). If the friend node is not within the communication range of the AP, it invokes other friend nodes in a recursive manner until the AP is reached. The concept of proxy RADIUS (Rigney et al., 2000) is used for ensuring forwarding compatibility and secure message exchange over multi-hops. Proxy chaining (Aboba & Vollbrecht, 1999) takes place if the friend node is not directly connected to an AP. To achieve higher level of security on each authenticated link between the communicating nodes, 802.11i encryption is used by invoking the four-way handshake between each MC and its authenticator (AP or friend node). This approach is useful in open mesh network scenarios, since it allows *authentication by delegation* among the mesh nodes. In addition, since the authentication keys are stored in the immediate nodes, the re-authentication process is optimized in case of roaming of the MCs. However, an adaptation is needed that allows establishment of multiple simultaneous connections to the authenticators - APs and the friend nodes – in a dense mesh topology. Also, a solution is needed to support fast and secure roaming across multiple *wireless mesh routers* (WMRs). A possible solution is through sharing session keys of authenticated clients among the WMRs (Moustafa, 2007).

## 4.3 Data packet authentication

An approach to prevent unauthorized node getting access to the network services in WMNs is to authenticate the transmitted data packets. Following this approach, a *lightweight hop-by-hop access protocol* (LHAP) (Zhu et al., 2003; Zhu et al., 2006) has been proposed for authenticating MCs for preventing resource consumption attacks in WMNs. LHAP implements light-weight hop-by-hop authentication, where intermediate nodes authenticate all the packets they receive before forwarding them further in the network. In this protocol, an MC first performs some light-weight authentication operations to bootstrap a trust relationship with its neighbors. It then invokes a light-weight protocol for subsequent traffic authentication and data encryption. LHAP is ideally suited for ad hoc networks, where it resides between the data link layer and the network layer and can be seamlessly integrated with secure routing protocols to provide high-level of security in a communication network.

LHAP employs a packet authentication technique based on the use of *one-way hash chains* (Lamport, 1981). Moreover, it uses TESLA (Perrig et al., 2001) protocol to reduce the number of public key operations for bootstrapping and maintaining trust among the nodes. For every traffic packet received from the network layer, LHAP adds its own header, which includes the node ID, a packet type field indicating a traffic packet, and an authentication tag. The packet is then passed to the data link layer and control packets are generated for establishing and maintaining trust relationships with the neighbor nodes. For a received packet, LHAP verifies its authenticity based on the authentication tag in the packet header. If the packet is valid, LHAP removes the LHAP header and passes the packet to the network layer; otherwise, it discards the packet. LHAP control packets are passed to the network layer with the goal to allow LHAP execution without affecting the operation of the other layers.

LHAP is very suitable for WMN applications. For secure roaming, LHAP can be useful in distributing session keys among MCs employing a special type of packet designated for this purpose. However, the focus of this protocol is on preventing resource consumption attack on the network. However, LHAP cannot prevent insider attacks and hence complementary mechanisms are needed for this purpose (Moustafa, 2007).

## 4.4 Proactive authentication and pre-authentication schemes

In (Pack & Choi, 2004), a fast handoff scheme based on prediction of mobility pattern has been proposed. In this scheme, an MC on entering in the coverage area of an access point performs authentication procedures for multiple MRs (or APs). When an MC sends an authentication request, the AAA server authenticates the all the relevant APs (or MRs) and sends multiple session keys to the MC. A prediction method known as *frequent handoff region* (FHR) selection is utilized to reduce the handoff delay further. FHR selection algorithm takes into account user mobility pattern, service classes etc. to make a selection of frequent MRs suitable for handoff. To increase the accuracy of the user mobility prediction, a proactive key distribution approach has been proposed in (Mishra et al., 2004). A new data structure – neighbor graphs – is used to determine the candidate MR sets for the MC to associate with.

A reliable re-authentication scheme has been proposed in (Aura & Roe, 2005), in which an MR issues a credential for the MC it is currently serving. The credential can be used later (by the next MR) to certify the authenticity of the MC.

A fast authentication and key exchange mechanism to support seamless handoff has been proposed in (Soltwisch et al., 2004). The mechanism uses the *context transfer protocol* (CTP) (Loughney et al., 2005) to forward session key from the previous router to the new access router.

### 4.5 Extensible authentication protocols

IEEE 802.1X has been applied to resolve some of the security problems in the 802.11 standard, where the MC and the AS authenticate each other by applying an upper layer authentication protocol like *extensible authentication protocol encapsulating transport layer security* (EAP-TLS) protocol (Aboba & Simon, 1999). Although EAP-TLS offers mutual authentication, it introduces high latency in WMNs because each terminal acts as an authenticator for its neighbor to reach the AS. This can lead to longer paths to the AS. Furthermore, in case of high mobility of terminals, re-authentication due to frequent handoffs can make be detrimental to real-time applications. Consequently, variants of EAP have been proposed by researchers to adapt 802.1X authentication model to multi-hop communications in WMNs. Some of these mechanisms are briefly discussed below.

**EAP with token-based re-authentication:** a fast and secure hand-off protocol is presented in (Fantacci et al., 2006), which allows mutual authentication and access control thereby preventing insider attacks during the re-authentication process. To achieve this, old authentication keys are revoked. Thus, a node should ask for the keys from its neighbors or from the AS when its needs the keys. The mechanism involves a token-based re-authentication scheme based on a two-way handshake between the node that performs the handshake and the AS. The AS is involved in every hand-off to have a centralized entity for monitoring the network. An authentication token, in the form of keying material is provided by the authenticator of the network to the AS to obtain the PMK key. The authenticator can be an AP or a host in the WMN. Initially, the MC performs a full EAP-TLS authentication, generating a PMK key that is then shared between the MC and its authenticator. Whenever the MC performs hand-off to another authenticator, the new authenticator should receive the PMK key to avoid a full re-authentication. The new authenticator issues a request to the AS for the PMK and adds a token to the request. The token is a cryptographic material to prove that the authenticator is in contact with the MC which owns the requested PMK. The token was earlier generated by the MC while performing the hand-off and was transmitted to the new authenticator. The AS verifies the token, and issues the PMK to the new authenticator. This protocol is secure and involves centralized key management. However, the need to involve the AS in each re-authentication is not suitable for scenarios where MCs have random and frequent mobility (Moustafa, 2007). A distributed token verification will be more suitable for open and multi-hop WMN environments.

**EAP-TLS over PANA:** a security architecture suitable for multi-hop mesh network is presented in (Cheikhrouhou et al., 2006) that employs EAP-TLS over *protocol for carrying authentication and network access* (PANA) (Parthasarathy, 2006). It proposes an authentication solution for WMNs adapting IEE 802.1X so that MCs can be authenticated by MRs. The authentication between MCs and MRs requires MCs to be directly connected to the MRs. Since PANA enables MCs to authenticate to the access network using IP protocol, it is used in this mechanism to overcome the problem of association between MCs and MRs that can be attached through more than one intermediate node. When a new MC joins the network, it first gets an IP address (pre-PANA address) from a local DHCP server. Then, the PANA protocol is initiated so that the mobile node discovers the *PANA access* (PAA) router to authenticate itself. After successful authentication, the MC initiates the *Internet key exchange* (IKE) protocol with the MR for establishing a security association. Finally, IPSec tunnel ensures data protection over the radio link and a data access control by the MR. During the authentication and authorization phases, PANA uses EAP message exchange between the MC and the PAA, where PAA relays EAP messages to the AS using EAP over RADIUS. EAP-TLS message is used in this approach. The protocol is suited for heterogeneous WMNs since it is independent of the technology of the wireless media. However, PANA requires use of IP addresses in the mesh nodes. This puts a restriction in its use since all elements of a WMN may not use IP as the addressing standard.

**EAP-TLS using proxy chaining:** the combinations of (Moustafa et al., 2006a; Moustafa et al., 2006b) propose adaptive EAP solutions for authentication and access control in the multi-hop wireless environment. In (Moustafa et al., 2006a), an adapted EAP-TLS approach is used to allow authentication of mobile nodes. A delegation process is used among mobile nodes by use of auxiliary authenticators in a recursive manner until the AS is reached. To allow extended forwarding and exchange of EAP-TLS authentication messages, proxy RADIUS is involved using proxy chaining among the intermediate nodes between the MCs requesting the authentication and the AS. This approach permits the storage of authentication keys of the MCs in the auxiliary authenticators. This speeds up the re-authentication process and enhances the performance of the adaptive EAP-TLS mechanism. This solution is applicable for WMNs, especially in multi-hop communications. However, to support secure roaming across different *wireless mesh routers* (WMRs), communication is required between the old and the new WMRs. This can be done by using central elements or switches that link the WMRs and allow storing of information in a central location and distribution of information among the WMRs.

**EAP-enhanced pre-authentication:** an EAP-enhanced pre-authentication scheme for mobile WMN (IEEE 802.e) in the link layer has been proposed in (Hur et al., 2008). In this scheme, the PKMv2 (public key management version 2) has been slightly modified based on the key hierarchy in a way that the communication key can be established between the MC and the target MR before hand-off in a proactive way. The modification allows the master session key generated by the authentication server to bind the MR identification (i.e., base station identification) and the MAC address of the MC. In the pre-authentication phase, the authentication server generates and delivers the unique public session keys for the neighbor MRs of the MC. The neighboring MRs are the access points that the MC potentially moves to. These MRs can use the public session key to derive an authorization key of the corresponding MC. In the same way, the MC can derive the public session key and the authorization key for its neighbor MRs, with the MR identification. Once the handoff is complete, the MC only needs to perform a three-way handshake and update the encryption key since the MC and MR already possess the authentication key. Thus a re-authentication with the authentication server is avoided and the associated delay is reduced.

**Distributed authentication:** a distributed authentication for minimizing the authentication delay has been proposed in (Lee et al., 2008), in which multiple trusted nodes are distributed over a WMN to act on the behalf of an authentication server. This makes management of the network easy, and it also incurs less storage overhead in the MRs. However, the performance of the scheme will degrade when multiple MCs send out their authentication requests, since the number of trusted nodes acting as the authentication server is limited compared to the number of access routers. In (He et al., 2010), a distributed *authenticated key establishment scheme* (AKES) has been proposed based on *hierarchical multi-variable symmetric functions* (HMSF). In this scheme, MCs and MRs can mutually authenticate each other and establish pair-wise communication keys without the need of interaction with any central authentication server. The authors have extended the polynomial-based key generation concept (Blundo et al., 1993) to the asymmetric function for mutual authentication among the MCs and MRs. Based on the symmetric polynomial and an asymmetric function, an efficient and hierarchical key establishment scheme is designed This substantially reduces the communication overhead and authentication delay.

**Secure authentication:** an improved security protocol for WMNs has been proposed in (Lukas & Fackroth, 2009). The protocol is named "WMNSec", which is based on the four-way handshake mechanism in 802.11i. In WMNSec, a dedicated station - *mesh key distributor* (MKD) – generates one single dynamically generated key for the whole network. This key is called the *global key* (GK). The GK is distributed from the MKD to the authenticated stations (MRs) using the four-way handshake from 802.11i. A newly joined MR would become another authenticator after it is authenticated and become the authenticated part of the WMN. Thus, the iterative authentication forms a spanning tree rooted as the MKD and spanning the whole network. To provide a high level of security, each key has a limited validity period. Periodic re-keying ensures that the keys used in all stations are up-to-date.

### 4.6 Authentication using identity-based cryptography

*Identity-based cryptography* (IBC) is a public key cryptography in which public key of a user is derived from some publicly available unique identity information about the user, e.g. SSN, email address etc. Although the concept of IBC was first introduced by Shamir (Shamir, 1984), a fully functional IBC scheme was not established till Boneh and Franklin applied Weil pairing to construct a bilinear map (Boneh & Franklin, 2001). Using IBC, an attack-resilient security architecture called "ARSA" for WMNs has been proposed in (Zhang & Fang, 2006). The relationship among three entities in this scheme, e.g., brokers, users and network operators are made analogous to that among a bank, a credit card holder, and a merchant. The broker acts as a TTP that distributes secure pass to each authenticated user. Each secure pass has the ID of the user enveloped in it and the WMN operator grants access to all the users those possess secure passes. The users are not bound to any specific operator, and can get ubiquitous network access by a universal pass issued by a *third-party broker*. ARSA also provides an efficient mutual *authentication and key agreement* (AKA) between a user and a serving WMN domain or between users served by the same WMN domain.

## 4.7 Privacy protection schemes in WMNs

Traffic privacy preservation is an important issue in WMNs. In a community mesh network, the traffic of mobile users can be observed by the MRs residing at its neighbors, which could reveal sensitive personal information. A mesh network privacy-preserving architecture is presented in (T. Wu et al., 2006). The mechanism aims to achieve traffic confidentiality based on the concept of *traffic pattern* concealment by controlling the routing process using multi-paths. The traffic from the source (i.e., IGW) to the destination (i.e., MR) is split into multiple paths. Hence, each relaying nodes along the path from the source to the destination can observe only a portion of the entire traffic. The traffic is split in a random manner (both spatially and temporally) so that an intermediate node can have little knowledge to figure out the overall traffic pattern. In this way the traffic confidentially is achieved. The mechanism defines an information-theoretic metric, and then proposes a penalty-based routing algorithm to allow traffic pattern hiding by exploiting multiple available paths between a pair of nodes. Source routing strategy is adopted so that a node can easily know the topology of its neighborhood. The protocol can also ensure communication privacy in WMNs, where each destination node is able to consistently limit the proportion of mutual information it shares with the observing node. However, the traffic splitting can increase delay in communication and hence this mechanism may not be suitable for real-time applications in WMNs.

A novel privacy and security scheme named PEACE (Privacy Enhanced yet Accountable seCurity framEwork) for WMNs has been proposed in (Ren et al., 2010). The scheme achieves explicit mutual authentication and key establishment between users (i.e. MCs) and MRs and between the users themselves (i.e., between the MCs). It also enables unilateral anonymous authentication between users and the MRs and bilateral anonymous authentication between a pair of users. Moreover, it enables user accountability by regulating user behaviors and protects WMNs from being abused and attacked. Network communications can be audited in cases of disputes and frauds. The high level architecture of PEACE trust model consists of four kinds of network entities: the network operator, user group managers, user groups and a *trusted third party* (TTP). Before accessing the WMN services, each user has to enroll in at least one user group whose manager, thus, knows the essential and non-essential attributes of the user. The users do not directly register with the network operator; instead, each group manager subscribes to the network operator on behalf of its group members. Upon registration from a group manager, the network operator allocates a set of group secret keys to this user group. The network operator divides each group secret key into two parts – one part is sent to the requesting group manager and the other part to the TTP. To access network services, each user request one part of the group secret key from his group manager and the other part from the TTP to recover a complete group secret key. The user also needs to return signed acknowledgments to both the group manager and the TTP. PEACE uses a variation of the short group signature scheme proposed in (Boneh & Shacham, 2004) to ensure sophisticated user privacy. The scheme is resistant to bogus data injection attacks, data phishing attacks and DoS attacks (Ren et al., 2010).

A security architecture named "SAT" has been proposed in (Sun et al., 2008; Sun et al., 2011). The system consists of ticket-based protocols, which resolves the conflicting security requirements of unconditional anonymity for honest users and traceability of misbehaving users in a WMN. By utilizing the tickets, self-generated pseudonyms, and the hierarchical identity-based cryptography, the architecture has been demonstrated to achieve the desired security objectives and the performance efficiency. The system uses a blind signature technique from the payment systems. (Brands, 1993; Wei et al., 2006; Figueiredo et al., 2005; Chaum, 1982), and hence it achieves the anonymity by delinking user identities from their activities. The pseudonym technique also renders user location information unexposed. The pseudonym generation mechanism does not rely on a central authority, e.g. the *broker* in (Zhang & Fang, 2006), the *domain authority* in (Ateniese et al., 1999), the *transportation authority* or the *manufacturer* in (Raya & Hubaux, 2007), and the *trusted authority* in (Zhang et al., 2006), who can derive the user's identity from his pseudonyms and illegally trace on an honest user. However, the system is not intended for achieving routing anonymity. *Hierarchical identity-based cryptography* (HIBC) for inter-domain authentication is adopted to avoid domain parameter certification in order to ensure anonymous access control.

## 5. The hierarchical architecture of a WMN

In this section, we first present a standard architecture of a typical WMN for which we propose a security and privacy protocol. The architecture is a very generic one that represents majority of the real-world deployment scenarios for WMNs. The architecture of a hierarchical WMN consists of three layers as shown in Fig. 5. At the top layers are the *Internet gateways* (IGWs) that are connected to the wired Internet. They form the backbone infrastructure for providing Internet connectivity to the elements in the second level. The entities at the second level are called wireless *mesh routers* (MRs) that eliminate the need for wired infrastructure at every MR and forward their traffic in a multi-hop fashion towards the IGW. At the lowest level are the *mesh clients* (MCs) which are the wireless devices of the users. Internet connectivity and peer-to-peer communications inside the mesh are two important applications for a WMN. Therefore design of an efficient and low-overhead communication protocol which ensure security and privacy of the users is a critical requirement which poses significant research challenges.

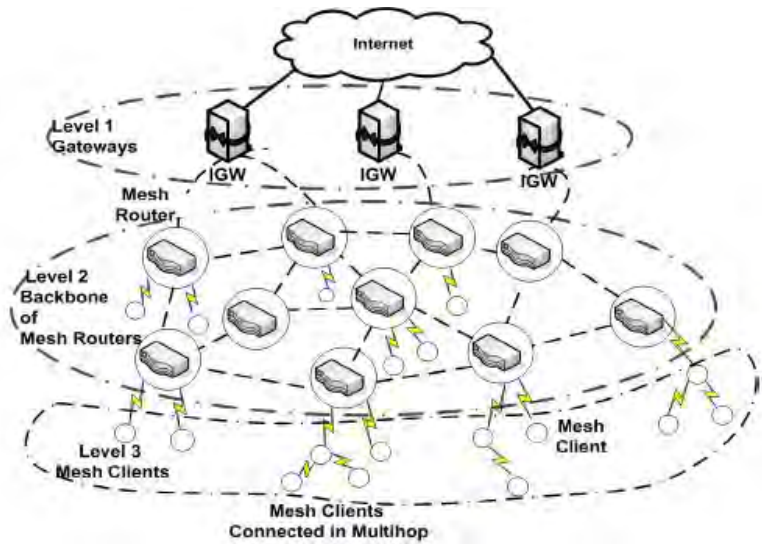

Fig. 5. A three-tier architecture of a wireless mesh network (WMN)

For designing the proposed protocol and to specify the WMN deployment scenario, the following assumptions are made.

1.  Each MR which is authorized to join the wireless backbone (through the IGWs), has two certificates to prove its identity. One certificate is used during the authentication phase that occurs when a new node joins the network. EAP-TLS (Aboba et al., 2004) for 802.1X authentication is used for this purpose since it is the strongest authentication method provided by EAP (Aboba et al., 2004), whereas the second certificate is used for the authentication with the *authentication server* (AS).
2.  The certificates used for authentication with the RADIUS server and the AS are signed by the same *certificate authority* (CA). Only recognized MRs are authorized to join the backbone.
3.  Synchronization of all MRs is achieved by use of the *network time protocol* (NTP) protocol (Mills, 1992).

The proposed security protocol serves the dual purpose of providing security in the access network (i.e., between the MCs and the MRs) and the backbone network (i.e., between the MRs and the IGWs). These are described the following sub-sections.

## 5.1 Access network security

The access mechanism to the WMN is assumed to be the same as that of a *local area network* (LAN), where mobile devices authenticate themselves and connect to an *access point* (AP). This allows the users to the access the services of the WMN exploiting the authentication and authorization mechanisms without installing any additional software. It is evident that such security solution provides protection to the wireless links between the MCs and the MRs. A separate security infrastructure is needed for the links in the backbone networks. This is discussed in Section 5.2.

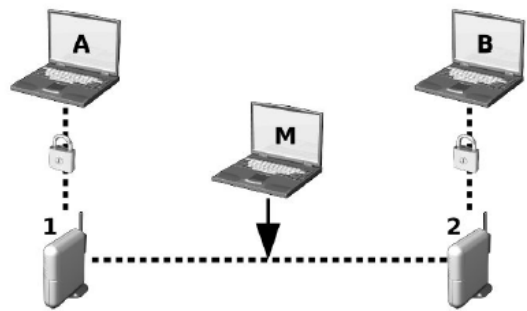

Fig. 6. Secure information exchange among the MCs *A* and *B* through the MRs 1 and 2

Fig. 6 illustrates a scenario where users *A* and *B* are communicating in a secure way to MRs 1 and 2 respectively. If the wireless links are not protected, an intruder *M* will be able to eavesdrop on and possibly manipulate the information being exchanged over the network. This situation is prevented in the proposed security scheme which encrypts all the traffic transmitted on the wireless link using a stream cipher in the data link layer of the protocol stack.

## 5.2 Backbone network security

For providing security for the traffic in the backbone network, a two-step approach is adopted. When a new MR joins the network, it first presents itself as an MC and completes the association formalities. It subsequently upgrades its association by successfully authenticating to the AS. In order to make such authentication process efficient in a high mobility scenario, the key management and distribution processes have been designed in a way so as to minimize the effect of the authentication overhead on the network performance. The overview of the protocol is discussed as follows.

Fig. 7 shows the three phases of the authentication process that a MR (say *N*) undergoes. When *N* wants to join the network, it scans all the radio channels to detect any MR that is already connected to the wireless backbone. Once such an MR (say *A*) is detected, *N* requests *A* for access to network services including authentication and key distribution. After connecting to *A*, *N* can perform the tasks prescribed in the IEEE 802.11i protocol to complete a mutual authentication with the network and establish a security association with the entity to which it is physically connected. This completes the Phase I of the authentication process. Essentially, during this phase, a new MR performs all the steps that an MC has to perform to establish a secure channel with an MR for authentication and secure communication over the WMN.

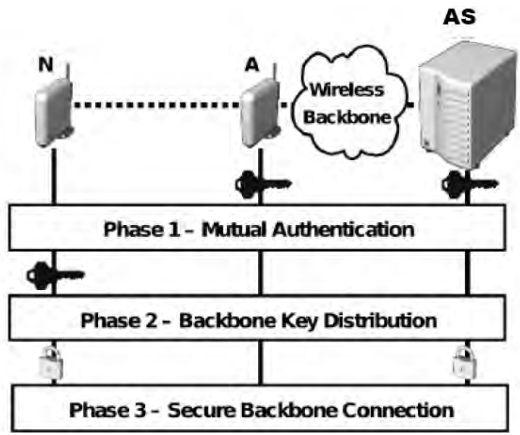

Fig. 7. Steps performed by a new MR (*N*) using backbone encrypted traffic to join the WMN

During Phase II of the authentication process, the MRs use the *transport layer security* (TLS) protocol. Only authorized MRs that have the requisite credentials can authenticate to the AS and obtain the cryptographic credentials needed to derive the key sequence used to protect the wireless backbone. In the proposed protocol, an end-to-end secure channel between the AS and the MR is established at the end of a successful authentication through which the cryptographic credentials can be exchanged in a secure way.

To eliminate any possibility of the same key being used over a long time, a server-initiated protocol is proposed for secure key management. The protocol is presented in Section 6. As mentioned earlier in this section, all the MRs are assumed to be synchronized with a central server using the NTP protocol.

Fig. 8 shows a collection of four MRs connected with each other by five wireless links. The MR *A* is connected with the AS by a wired link. At the time of network bootstrapping, only node *A* can connect to the network as an MR, since it is the only node that can successfully authenticate to the AS. Nodes *B* and *C* which are neighbors of *A* then detect a wireless network to which can connect and perform the authentication process following the IEEE 802.11i protocol. At this point of time, nodes *B* and *C* are successfully authenticated as MCs. After their authentication as MCs, nodes *B* and *C* are allowed to authenticate to the AS and request the information used by *A* to produce the currently used cryptographic key for communication in the network. After having derived such key, both *B* and *C* will be able to communicate with each other, as well as with node *A*, using the ad hoc mode of communication in the WMN. At this stage, *B* and *C* both have full MR functionalities. They will be able to turn on their access interface for providing node *D* a connection to the AS for joining the network.

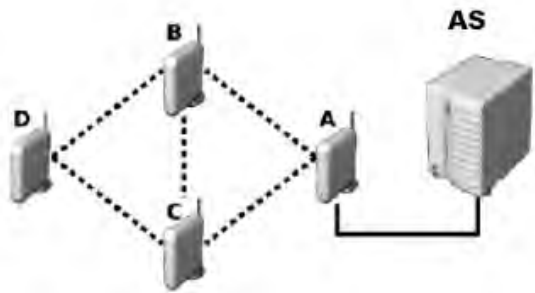

Fig. 8. Autonomous configuration of the MRs in the proposed security scheme

## 6. The key distribution protocol

In this section, the details of the proposed key distribution and management protocol are presented. The protocol is essentially a server-initiated protocol (Martignon et al., 2008) and provides the clients (MRs and MCs) flexibility and autonomy during the key generation.

In the proposed key management protocol delivers the keys to all the MRs from the AS in a reactive manner. The keys are used subsequently by the MRs for a specific time interval in their message communications to ensure integrity and confidentiality of the messages. After the expiry of the time interval for validity of the keys, the existing keys are revoked and new keys are generated by the AS. Fig. 9 depicts the message exchanges between the MRs and the AS during the execution of the protocol.

A newly joined MR, after its successful mutual authentication with a central server, sends its first request for key list (and its time of generation) currently being used by other existing MRs in the wireless backbone. Let us denote the *key list timestamp* as $TS_{KL}$. Let us define a *session* as the maximum time interval for validity of the key list currently being used by each node MR and MC). We also define the duration of a session as the product of the *cardinality of the key list* (i.e., the number of the keys in the key list) and the longest time interval of validity of a key (the parameter *timeout* in Fig. 9).

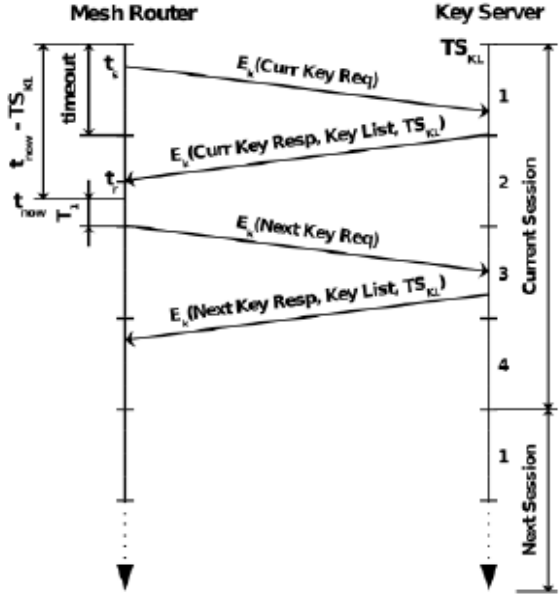

Fig. 9. Message exchanges between an MR and the AS in the key management protocol

The validity of a key list is computed from the time instance when the list is generated (i.e., $TS_{KL}$) by the AS. An MR, based on the time instance at which it joins the backbone ($t_{now}$ in Fig. 9), can find out the key (from the current list) being used by its peers ($key_{idx}$) and the interval of validity of the key ($T_i$) using (1) and (2) as follows:

$$key_{idx} = \left\lfloor \frac{t_{now} - TS_{KL}}{timeout} \right\rfloor + 1 \tag{1}$$

$$T_i = key_{idx} * timeout - (t_{now} - TS_{KL}) \tag{2}$$

In the proposed protocol, each WMN node requests the AS for the *key list* that will be used in the next session before the expiry of the current session. This is feature is essential for nodes which are located multiple hops away from the AS, since, responses from the AS take longer time to reach these nodes. The responses may also get delayed due to fading or congestion in the wireless links. If the nodes send their requests for key list to the AS just before expiry of the current session, then due to limited time in hand, only the nodes which have good quality links with the AS will receive the key list. Hence, the nodes which will fail to receive responses for the server will not be able to communicate in the next session due to non-availability of the current key list. This will lead to an undesirable situation of network partitioning.

The *key index* value that triggers the request from the nodes to the server can be set equal to the difference between the *cardinality of the list* and a *correction factor*. The correction factor can be estimated based on parameters like the network load, the distance of the node from the AS and the time required for the previous response.

In the proposed protocol, the correction factor is estimated based on the time to receive the response from the AS using (3), where $t_s$ is the time instance when the first key request was sent, $t_r$ is the time instance when the key response was received from the AS, and *timeout* is the validity period of the key. Therefore, if a node fails to receive a response (i.e., the key list) from the AS during timeout, and takes a time $t_{last}$, it must send the next request to the AS before setting the last key.

$$c = \left\lceil \frac{t_{last} - timeout}{timeout} \right\rceil \text{ if } t_{last} \geq timeout \qquad (3)$$

$$= 0 \text{ if } t_{last} < timeout$$

$$t_{last} = t_r - t_s$$

The first request of the key list sent by the new node to the AS is forwarded by the peer to which it is connected as an MC through the wireless access network. However, the subsequent requests are sent directly over the wireless backbone.

## 7. The privacy and anonymity protocol

As mentioned in Section 1, to ensure privacy of the users, the proposed security protocol is complemented with a privacy protocol so as to ensure user anonymity and privacy. The same *authentication server* (AS) used in the security protocol is used for managing the key distribution for preserving the privacy. To enable user authentication and anonymity, a novel protocol has been designed extending the *ring signature authentication* scheme in (Cao et al., 2004). It is assumed that a symmetric encryption algorithm $E$ exists such that for any key $k$, the function $E_k$ is a permutation over $b$-bit strings. We also assume the existence of a family of *keyed combining functions* $C_{k,v}(y_1, y_2, ...., y_n)$, and a publicly defined *collision-resistant* hash function $H(.)$ that maps arbitrary inputs to strings of constant length which are used as keys for $C_{k,v}(y_1, y_2, ...., y_n)$ (Rivest et al., 2001). Every keyed combining function $C_{k,v}(y_1, y_2, ...., y_n)$ takes as input the key $k$, an initialization $b$-bit value $v$, and arbitrary values $y_1, y_2, ...., y_n$. A user $U_i$ who wants to generate a session key with the authentication server, uses a ring of $n$ logged-on-users and performs the following steps.

**Step 1.** $U_i$ chooses the following parameters: (i) a large prime $p_i$ such that it is hard to compute discrete logarithms in $GF(p_i)$, (ii) another large prime $q_i$ such that $q_i \mid p_i - 1$, and (iii) a generator $g_i$ in $GF(p_i)$ with order $q_i$.

**Step 2.** $U_i$ chooses $x_{A_i} \in Z_{q_i}$ as his private key, and computes the public key $y_{A_i} = g_i^{x_{A_i}} \bmod p_i$.

**Step 3.** $U_i$ defines a trap-door function $f_i(\alpha, \beta) = \alpha . y_{A_i}^{\alpha \bmod q_i} . g_i^{\beta} \bmod p_i$. Its inverse function $f_i^{-1}(y)$ is defined as $f_i^{-1}(y) = (\alpha, \beta)$, where $\alpha$ and $\beta$ are computed as follows ($K$ is a random integer in $Z_{qi}$.

$$\alpha = y_{Ai} \cdot g_i^{-K.(g_i^K \bmod p_i) \bmod q_i} \bmod p_i \tag{4}$$

$$\alpha^* = \alpha \bmod q_i \tag{5}$$

$$\beta = K.(g_i^K \bmod p_i) - x_{Ai}.\alpha^* \bmod q_i \tag{6}$$

$U_i$ makes $p_i$, $q_i$, $g_i$ and $y_{A_i}$ public, and keeps $x_{A_i}$ as secret.

The *authentication server* (*AS*) chooses: (i) a large prime $p$ such that it is hard to compute discrete logarithms in $GF(p)$, (ii) another large prime $q$ such that $q \mid p - 1$, (iii) a generator $g$ in $GF(p)$ with order $q$, (iv) a random integer $x_B$ from $Z_q$ as its private key. *AS* computes its public key $y_B = g^{x_B} \bmod p$ and publishes $(y_B, p, q, g)$.

***Anonymous authenticated key exchange***: The key-exchange is initiated by the user $U_i$ and involves three rounds to compute a secret session key between $U_i$ and *AS*. The operations in these three rounds are as follows:

*Round 1*: When $U_i$ wants to generate a session key on the behalf of $n$ ring users $U_1$, $U_2$, ……$U_n$, where $1 \le i \le n$, $U_i$ does the following:

i.   (i) $U_i$ chooses two random integers $x_1$, $x_A \in Z_q^*$ and computes the following: $R = g^{x_1} \bmod p$, $Q = y_B^{x_1} \bmod p \bmod q$, $X = g^{x_A} \bmod p$ and $l = H(X, Q, V, y_B, I)$.

ii.  (ii) $U_i$ Chooses a pair of values $(\alpha_t, \beta_t)$ for every other ring member $U_t$ ($1 \le t \le n, t \ne k$) in a pseudorandom way, and computes $y_t = f_t(\alpha_t, \beta_t) \bmod p_t$.

iii. (iii) $U_i$ randomly chooses a *b*-bit initialization value $v$, and finds the value of $y_i$ from the equation $C_{k,v}(y_1, y_2, \ldots \ldots y_n) = v$.

iv.  (iv) $U_i$ computes $(\alpha_i, \beta_i) = f_i^{-1}(y_i)$ by using the trap-door information of $f_i$. First, it chooses a random integer $K \in Z_{q_i}$, computes $\alpha_i$ using (6), and keeps $K$ secret. It then computes $\alpha_i^*$ using (5) and finally computes $\beta_i$ using (6).

v.   (v) $(U_1, U_2, .. U_n, v, V, R, (\alpha_1, \beta_1), (\alpha_2, \beta_2), .., (\alpha_n, \beta_n)$ is the ring signature $\sigma$ on $X$.

Finally, $U_i$ sends $\sigma$ and $I$ to the server *AS*.

*Round 2*: *AS* does the following to recover and verify $X$ from the signature $\sigma$.

i.   *AS* computes $Q = R^{x_B} \bmod p \bmod q$, recovers $X$ using $X = V.g^Q \bmod p$ and hashes $X$, $Q$, $V$ and $y_b$ to recover $l$, where $l = H(X, Q, V, y_b, I)$.

ii.  *AS* computes $y_t = f_i(\alpha_t, \beta_t) \bmod p_i$, for $t$ = 1,2,…..$n$.

iii. *AS* checks whether $C_{k,v}(y_1, y_2, \ldots \ldots y_n) = v$. If it is true, *AS* accepts $X$ as valid; otherwise, *AS* rejects $X$. If $X$ is valid, *AS* chooses a random integer $x_b$ from $Z_q^*$, and computes the following: $Y = g^{x_b} \bmod p$   $K_s = X^{x_b} \bmod p$ and $h = H(K_s, X, Y, I')$. *AS* sends $\{h, Y, I'\}$ to $U_i$.

*Round 3*: $U_i$ verifies whether $K_S^{'}$ is from the server *AS*. For this purpose, $U_i$ computes $K_S^{'} = Y^{x_a} \bmod p$, hashes *K*, *X*, *Y* to get $h^{'}$ using $h^{'} = H(K_s^{'}, X, Y, I^{'})$. If $h^{'} = h$, $U_i$ accepts $K_s$ as the session key.

*Security analysis*: The key exchange scheme satisfies the following requirements.

*User anonymity*: For a given signature *X*, the server can only be convinced that the ring signature is actually produced by at least one of the possible users. If the actual user does not reveal the seed *K*, the server cannot determine the identity of the user. The strength of the anonymity depends on the security of the pseudorandom number generator. It is not possible to determine the identity of the actual user in a ring of size *n* with a probability greater than $1/n$. Since the values of *k* and *v* are fixed in a ring signature, there are $(2^b)^{n-1}$ number of $(x_1, x_2, ... x_n)$ that satisfy the equation $C_{k,v}(y_1, y_2, ... y_n) = v$, and the probability of generation of each $(x_1, x_2, ... x_n)$ is the same. Therefore, the signature can't leak the identity information of the user.

*Mutual authentication*: In the proposed scheme, not only the server verifies the users, but the users can also verify the server. Because of the hardness of inverting the hash function *f(.)*, it is computationally infeasible for the attacker to determine $(\alpha_i, \beta_i)$, and hence it is infeasible for him to forge a signature. If the attacker wants to masquerade as the *AS*, he needs to compute $h = H(K_s, X, Y)$. He requires $x_B$ in order to compute *X*. However, $x_B$ is the private key of *AS* to which the attacker has no access.

*Forward secrecy*: The forward secrecy of a scheme refers to its ability to defend leaking of its keys of previous sessions when an attacker is able to catch hold of the key of a particular session. The forward secrecy of a scheme enables it to prevent *replay attacks*. In the proposed scheme, since $x_a$ and $x_b$ are both selected randomly, the session key of each period has not relation to the other periods. Therefore, if the session key generated in the period *j* is leaked, the attacker cannot get any information of the session keys generated before the period *j*. The proposed protocol is, therefore, resistant to replay attack.

## 8. Performance evaluation

The proposed security and privacy protocols have been implemented in the Qualnet network simulator, version 4.5 (Network Simulator, Qualnet). The simulated network consists of 50 nodes randomly distributed in the simulation area forming a dense WMN. The WMN topology is shown in Fig. 10, in which 5 are MRs and remaining 45 are MCs. Each MR has 9 MCs associated with it. To evaluate the performance of the security protocol, first the network is set as a full-mesh topology, where each MR (and also MC) is directly connected to two of its neighbors. In such scenario, the throughput of a TCP connection established over a wireless link is measured with the security protocol activated in the nodes. The obtained results are then compared with the throughput obtained on the same wireless link protected by a static key to encrypt the traffic.

After having 10 simulation runs, the average throughput of a wireless link between a pair of MRs was found to be equal to 30.6 MBPS, when the link is protected by a static key. However, the average throughput for the same link was 28.4 MBPS when the link was protected by the proposed security protocol. The results confirm that the protocol does not cause any significant overhead on the performance of the wireless link, since the throughput in a link on average decreased by only 7%.

The impact of the security protocol for key generation and revocation on packet drop rate in real-time applications is also studied in the simulation. For this purpose, a VoIP application is invoked between two MRs which generated UDP traffic in the wireless link. The packet drop rates in wireless link when the link is protected with the proposed security protocol and when the link is protected with a static key. The transmission rate was set to 1 MBPS. The average packet drop rate in 10 simulation runs was found to be only 4%. The results clearly demonstrate that the proposed security scheme has no adverse impact on packet drop rate even if several key switching (regeneration and revocation) operations are carried out.

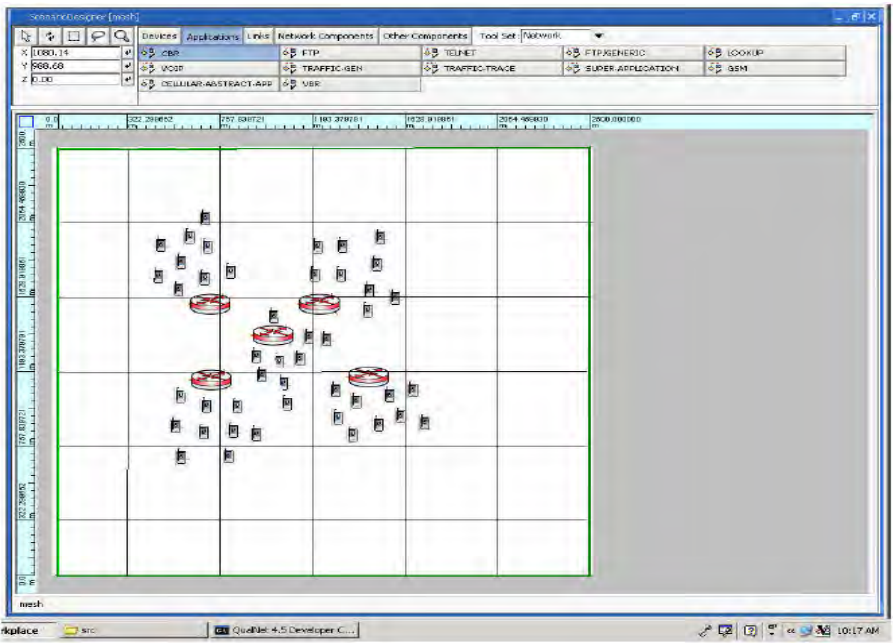

Fig. 10. The simulated network topology in Qualnet Simulator

The performance of the privacy protocol is also analyzed in terms of its storage, communication overhead. Both storage and communication overhead were found to increase linearly with the number of nodes in the network. In fact, it has been analytically shown that overhead due to cryptographic operation on each message is: $60n + 60$ bytes, where $n$ represents the number of public key pairs used to generate the ring signature (Xiong et al., 2010). It is clear that the privacy protocol has a low overhead.

## 9. Conclusion and future work

WMNs have become an important focus area of research in recent years owing to their great promise in realizing numerous next-generation wireless services. Driven by the demand for rich and high-speed content access, recent research has focused on developing high performance communication protocols, while security and privacy issues have received relatively little attention. However, given the wireless and multi-hop nature of communication, WMNs are subject to a wide range of security and privacy threats. This chapter has provided a comprehensive discussion on the current authentication, access control and user privacy protection schemes for WMNs. It has also presented a novel security and key management protocol that can be utilized for secure authentication in WMNs. The proposed security protocol ensures security in both the access and the backbone networks. A user privacy protection algorithm has also been presented that enables anonymous authentication of the users. Simulation results have shown the effectiveness of the protocol. Future research issues include the study of a distributed and collaborative system where the authentication service is provided by a dynamically selected set of MRs. The integration with the current centralized scheme would increase the robustness of the proposed protocol, maintaining a low overhead since MRs would use the distributed service only when the central server is not available. Authentication on the backbone network in a hybrid and open WMN is still an unsolved problem. In addition, authentication between MRs and IGWs from different operators in a hybrid WMN environment is another challenge. Authentication and key distribution in a mobile WMN such as mobile WiMAX or LTE networks is another open problem. High mobility users make the challenge even more difficult. Owing to very limited coverage IEEE 802.11-based MRs (e.g., 100 meters), the high-mobility users (e.g. a user on a fast moving car) will migrate from the coverage area of an MR to that of another. It is not acceptable for the user to authenticate and negotiate the key with each MR. Novel solutions possibly using group keys are needed for this purpose. The requirements of user anonymity and privacy of users should be integrated to most of the applications in WMNs.